\documentclass[final]{svjour3}
\usepackage{graphicx}
\usepackage{rotating}
\usepackage{amssymb}
\usepackage{mathptmx}
\usepackage[square,numbers]{natbib}
\usepackage [latin1]{inputenc}
\makeatletter
\journalname{Journal of Low Temperature Physics}

\begin{document}

\newcommand{\hdblarrow}{H\makebox[0.9ex][l]{$\downdownarrows$}-}
\title{QUBIC: the Q \& U Bolometric Interferometer for Cosmology}

\author{E.S. Battistelli$^{1, 2}$, P. Ade$^{3}$, J.G. Alberro$^{4}$, A. Almela$^{5}$, G. Amico$^{1}$, L. H. Arnaldi$^{6}$, D. Auguste$^{7}$, J. Aumont$^{8}$, S. Azzoni$^{9}$, S. Banfi$^{10, 11}$
% ...
}

\institute{$^{1}$ Universit\`a di Roma La Sapienza, Roma, Italy; 
$^{2}$ Istituto Nazionale di Fisica Nucleare Roma 1 Section, Roma, Italy;
$^{3}$ Cardiff University, Cardiff, UK;
$^{4}$ Univ. Nacional de la Plata, Argentina;
$^{5}$ Instituto de Tecnolog\'ias en Detecci\'on y Astropart\'iculas, Argentina;
$^{6}$ Ctr. At\'omico Bariloche e Instituto Balseiro, CNEA, Argentina;
$^{7}$ Laboratoire de l'Acc\'el\'erateur Lin\'eaire (CNRS-IN2P3), Orsay, France;
$^{8}$ Institut de Recherche en Astrophysique et Plan\'etologie (CNRS-INSU), Toulouse, France;
$^{9}$ University of Manchester, Manchester, UK;
$^{10}$ Universit\`a degli Studi di Milano Bicocca, Milano, Italy;
$^{11}$ Istituto Nazionale di Fisica Nucleare Milano Bicocca section, Milano, Italy;
$^{12}$ Centre de nanosciences et de nanotechnologies, Orsay, France;
$^{13}$ National University of Ireland, Maynooth, Ireland;
$^{14}$ Centre de Spectrom\'etrie Nucl\'eaire et de Spectrom\'etrie de Masse (CNRS-IN2P3), Orsay, France;
$^{15}$ University of Milan, Dept. of Physics, Milano, Italy;
$^{16}$ Astroparticule et Cosmologie (CNRS-IN2P3), Paris, France;
$^{17}$ Comisi\'on Nacional De Energia At\'omica, Argentina;
$^{18}$ Richmond University, Richmond, VA, USA;
$^{19}$ Universit\`a di Roma Tor Vergata, Roma, Italy;
$^{20}$ Universidad Nacional de San Martin, San Martin, Argentina;
$^{21}$ Institut d'Astrophysique Spatiale (CNRS-INSU), Orsay, France;
$^{22}$ Agenzia Spaziale Italiana, Rome, Italy;
$^{23}$ Pontificia Universidad Catolica de Chile, Santiago, Chile;
$^{24}$ Instituto Argentino de Radioastronomi­a, Argentina;
$^{25}$ Kavli Institute for Particle Astrophysics and Cosmology, Stanford, California, USA;
$^{26}$ California Institute of Technology, Pasadena, California, USA;
$^{27}$ Istituto Nazionale di Fisica Nucleare Pisa Section, Pisa, Italy;
$^{28}$ University of Wisconsin, Madison, WI, USA;
$^{29}$ Brown University, Providence, RI, USA;
$^{30}$ Regional Noroeste (CNEA), Argentina;
$^{31}$ Department of Physics, University of Surrey, Guildford, UK. \email{elia.battistelli@roma1.infn.it}}

\authorrunning{Battistelli et al} 
\titlerunning{QUBIC}
\maketitle

\begin{abstract}
The Q \& U Bolometric Interferometer for Cosmology, QUBIC, is an innovative experiment designed to measure the polarization of the Cosmic Microwave Background and in particular the signature left therein by the inflationary expansion of the Universe. The expected signal is extremely faint, thus extreme sensitivity and systematic control are necessary in order to attempt this measurement. QUBIC addresses these requirements using an innovative approach combining the sensitivity of Transition Edge Sensor cryogenic bolometers, with the deep control of systematics characteristic of interferometers. This makes QUBIC unique with respect to others classical imagers experiments devoted to the CMB polarization. In this contribution we report a description of the QUBIC instrument including recent achievements and the demonstration of the bolometric interferometry performed in lab. QUBIC will be deployed at the observation site  in Alto Chorrillos, in Argentina at the end of 2019.
\end{abstract}

\section{Introduction}
The Cosmic Microwave Background (CMB) is one of the major tools in cosmology. Measurements of the CMB anisotropies and the linear polarization originated by the density (scalar) fluctuations in the primordial plasma, are allowing cosmologists to delineate a cosmological Standard Model entering in an era of precision cosmology. Density perturbations generate a characteristic curl-free pattern of CMB polarization which has been observed by several experiments (see eg. \cite{planck16, bicep18}).
The search of the signature of an inflationary expansion of the Universe in its very early stage, is the next milestone in modern cosmology. This is possible by means of the detection of the curl component of the polarization of the CMB, B-modes, originated by tensor perturbations (a stochastic background of gravitational waves) generated during the predicted inflationary exponential expansion of the Universe occurred when it was only $\sim$10$^{-35}$s old. The energy scale (the potential V) at which inflation took place, can be described in terms of the ratio between the amplitude of tensor and scalar fluctuations, the tensor-to-scalar ratio $r$:
\begin{equation}
V^{1/4}\approx \left({\frac{r}{0.01}}\right)^{1/4} (10^{16}{\rm GeV})
\end{equation}
A value of $r$ close to 0.01 would imply an energy scale close to what is theoretically foreseen by grand unified theories of
fundamental interactions. We stress, however, that there is no firm prediction for the value of $r$ in inflationary models.
A detection of primordial B-modes of the polarization of the CMB (as opposed to gravitational lensing-induced B-modes), at angular scales ranging from several degrees to $\lesssim$1$^{\circ}$ would represent a footprint of this early inflationary phase. The expected signal is extremely weak and can be as low as a few nK or lower. Polarized foreground signals and instrumental systematics have to be extremely well characterized and controlled. So far, no experiments have been able to detect the primordial B-modes of the CMB polarization. The best upper limits are those set by the BICEP-Keck collaboration to a level of $r<$0.07 at 95\% C.L. \cite{bicep18}.

The Q \& U Bolometric Interferometer for Cosmology (QUBIC)  \cite{osall18, tar16, pia12, bat11} is designed to measure B-modes, or set upper limits down to a level of $r <$0.01. This will be done by observing the sky in two spectral bands centered at frequencies of 150GHz and 220GHz, for an efficient foreground removal, with spectroscopic capability within each observational band. QUBIC aims at controlling the systematic effects by acquiring the interferometric pattern of the observed sources, and performing the so called self-calibration during which single Fourier modes from the sky are measured redundantly \citep{self}. The required sensitivity is ensured by two arrays of 1024 pixels of Transition Edge Sensors (992 of which exposed to radiation) cooled down at $\sim$300mK. The sensitivity to polarization is ensured by the presence of a rotating Half-Wave Plate (HWP) followed by a polarizing wire grid.

\section{QUBIC instrument}
A drawing of the QUBIC cryostat is shown in Fig.~1. The signal from the sky is detected after having gone through several optical components, all located within the QUBIC cryostat. Radiation goes through a vacuum window, a rotating HWP, a polarizing wire grid, an array of 400 back-to-back corrugated horns, a dual mirror beam-combiner, a dichroic filter able to separate the two QUBIC observational bands, thermal and band-pass quasi-optical filters on each of the thermal stages of the cryostat, and then is detected by two arrays of TESs. Differently from classical imagers, in QUBIC radiation is blocked by the array of corrugated horns, after which a two mirrors beam combiner focuses radiation into the detectors array. The result is that QUBIC act as synthetic imager which detects the convolution between the sky brightness and the synthesized beam of each pixel of the instrument. 
\begin{figure}[htbp]
\begin{center}
\includegraphics[width=1.1\linewidth, keepaspectratio]{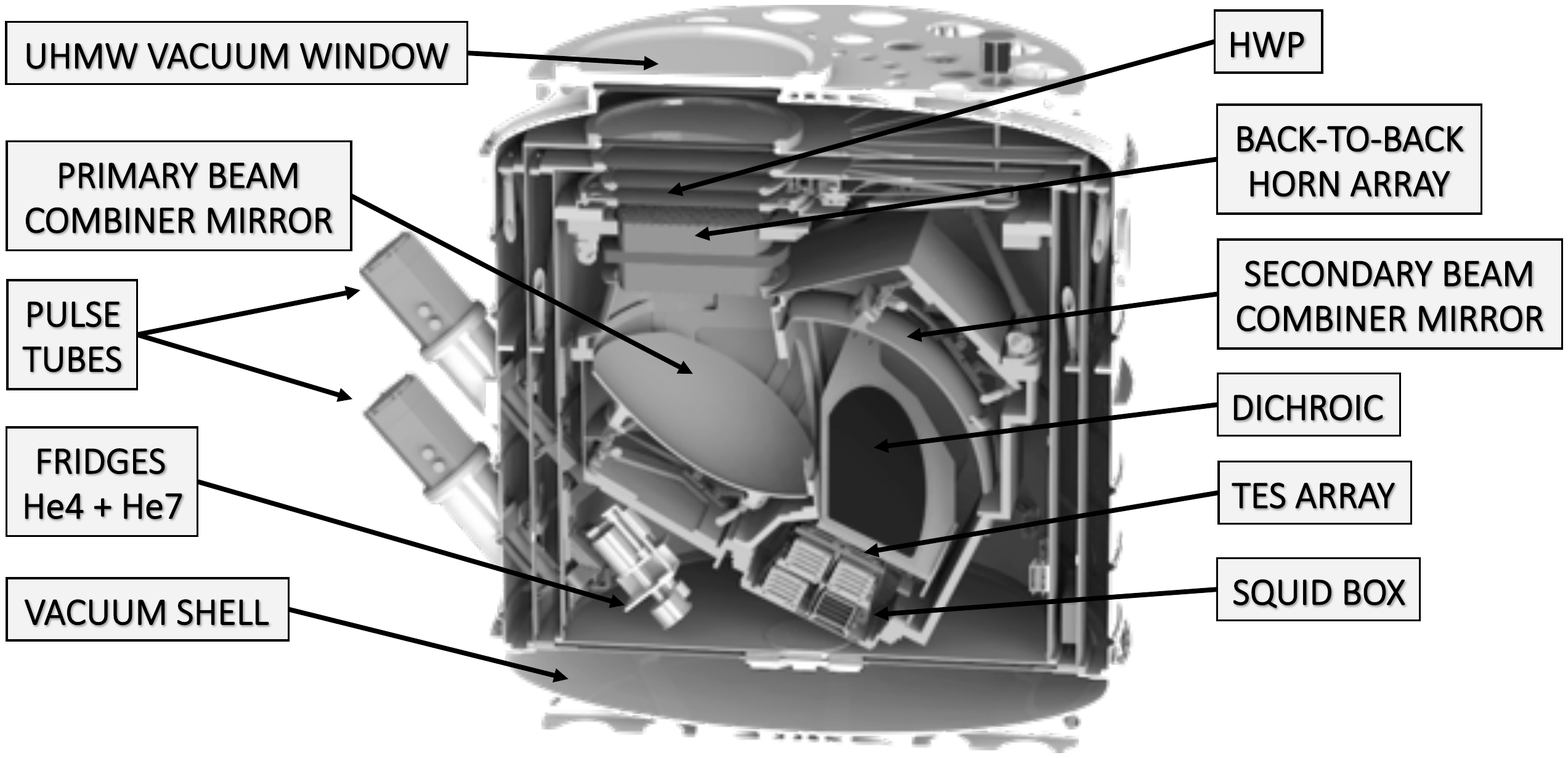}
\caption{3D rendering cut of the QUBIC cryostat with internal parts highlighted.}
\end{center}
\label{qubic2}
\end{figure}

\subsection{Cryogenics}
The cryogenic system of QUBIC is based on the use of a large (1.4m$\times$1.55m[h]) aluminum vacuum shell, with shape and structure optimized for withstanding the stress of atmospheric pressure outside the cryostat. Two pulse tube (PT) mechanical cryocoolers, allow us to cool a large volume ($\sim 1$m$^{3}$) at 4K with 0.9W cooling power each. The vacuum window is a 50cm diameter, 20mm thick Ultra High Molecular Weight (UHMW) polyethylene window. The two PT ensure a first thermal stage of $\sim$40K and a second stage at  $\sim$4K to which are connected radiation shields coated with multi layer insulation (MLI). A $^{4}$He closed cycle evaporation refrigerator allows us to cool to $\sim$1K an additional shield in which the beam-combiner and the dichroic filter are hosted (1K box). The 1K box is mounted on the 4K stage through a carbon fibre hexapod assembly to ensure mechanical robustness and thermal isolation. During cool-down, in order to reduce operation times, a combination of a mechanical heat switch and two convective heat switches are used until the 1K box temperature reaches 4K. A $^{3}$He$+$$^{4}$He ($^{7}$He) closed cycle evaporation refrigerators are used to cool down at 0.3K the detector arrays, including their pass-band filters and proximity read-out electronics \cite{may18}.

\subsection{Optics}
During its path toward the TES array of detectors, radiation encounters a set of 400 co-aligned primary horns. They represent the array of apertures of the interferometer. The horns are fabricated using the platelets technique (see e.g. \cite{del11}). Each aperture is a back-to-back double horn with a RF switch between the two, allowing us to block radiation on a single aperture basis and effectively to exclude single baselines. The back horns illuminate a double mirror beam-combiner after which a dichroic filter reflects radiation toward the 220GHz array and transmits radiation toward the 150GHz array of detectors.

The sensitivity to polarization is ensured by means of a rotating HWP, followed by a polarizing wire-grid: a Stokes polarimeter. The stepped rotation of the HWP is generated by an external stepper motor and is then transmitted on the top of the 4K stage, where the device is placed, through a magnetic joint, a fiberglass tube, and a set of pulleys. The signal $D$ measured at time $t$ by a detector $p$ on the focal plane is:
\begin{equation}
D(p,\nu,t)=G[S_{I}(p,\nu)+cos(4\phi_{HWP}(t))S_{Q}(p,\nu)+sin(4\phi_{HWP}(t))S_{U}(p,\nu)]
\end{equation}
where $\nu$ is the frequency, $\phi_{HWP}$ is the angle of the HWP at time $t$, and $G$ is an overall calibration constant that takes into account the efficiency of the optical chain. The three terms $S_{I,Q,U}$ represent the sky signal in intensity and polarization convolved with the so-called synthesized beam (see eq. \ref{eq3}). One of the key features of QUBIC is its spectral capability, which allows the synthesized beam falling on the detector arrays to be separated into sub-bands of the instrument's bandwidths. This effect can be used monitor the spectral dependence of different signals in the sky, and to mitigate non-idealities related to HWP performance over the bandwidth using a narrow band artificial source during the self-calibration.

\subsection{Detection chain}
Each focal plane is composed of four 256-pixels arrays of Nb$_{x}$Si$_{1-x}$ TESs with a critical temperature of $\sim$500mK. QUBIC TESs are optimized for a 5-50pW background suspending the absorbers and the thermistors on a 500nm thin SiN membrane, resulting in a thermal conductivity between 50 and 500pW/K. The total noise equivalent power (NEP) of these detectors is $\sim 5 \times 10^{-17}W/Hz^{0.5}$, with a time constant between 10 and 100ms. The two thermistor ends are routed to the edge of the wafer by means of superconducting Al lines, and connected via wire-bonds to the front-end electronics. The TESs are voltage-biased to exploit strong electrothermal feedback. The readout is based on SQ600S\footnote{https://starcryo.com/} SQUIDs which allow for 128:1 time-domain multiplexing, using cryogenic ASIC low noise amplifiers. Details of the fabrication of the TESs and of the entire detection chain of the QUBIC experiment can be found in these proceedings (\cite{pia19, mar19}).

\section{QUBIC as a Bolometric Interferometer} 
\begin{figure}[ht]
\begin{center}
\includegraphics[width=1.02\linewidth]{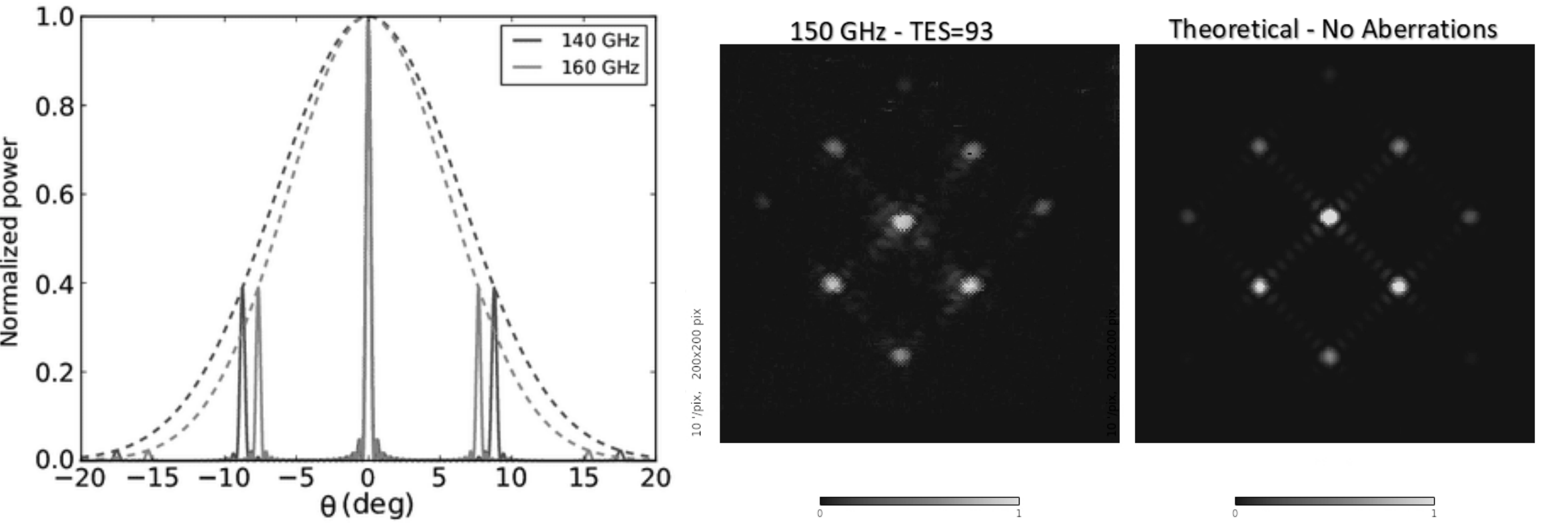}
\caption{\emph{Left}: Synthesized beam for the central pixel at 140 and 160 GHz. The separation between primary and secondary maxima depends on the frequency of the signal demonstrating the spectral capability of QUBIC within each spectral bandwidth. 
\emph{Center/Right}: Measured (\emph{centre}) and simulated (\emph{right}) synthesized beam of one of the detectors (93$^{rd}$) in the 150GHz array of the TD. Each map shows 200x200 pixels with 10'/pixel on the sky.}  
\label{fig:2}
\end{center}
\end{figure}

\begin{figure}[ht]
\begin{center}
\includegraphics[width=1.0\linewidth]{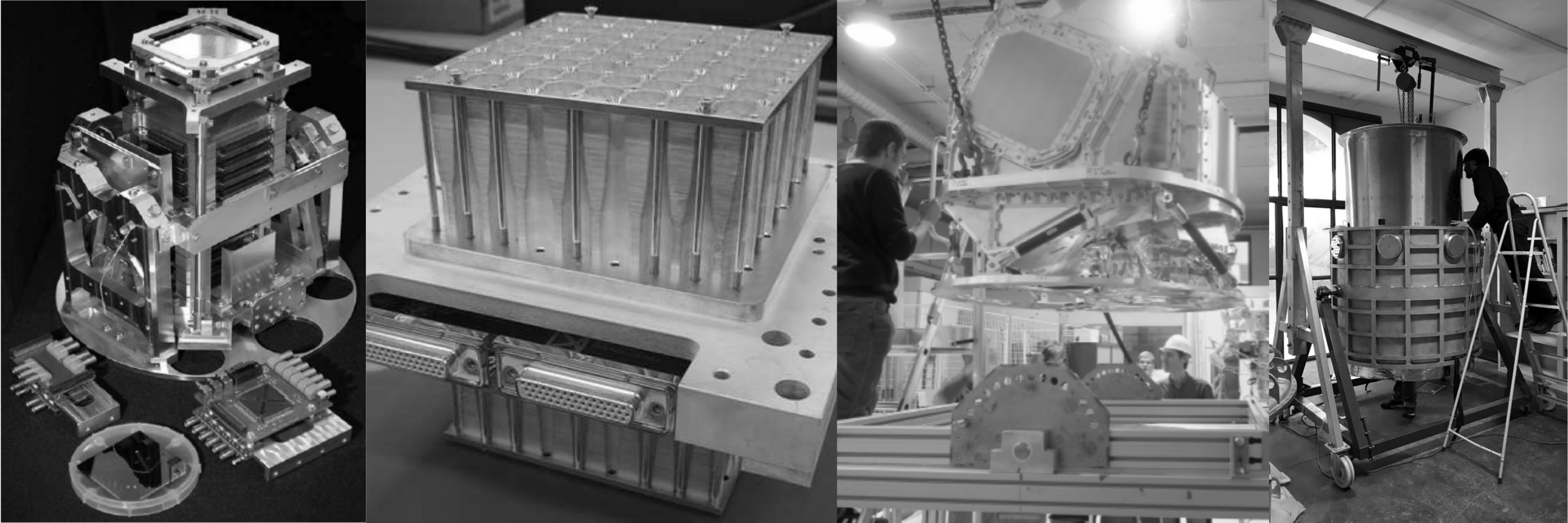}
\caption{From \emph{left} to \emph{right}: the cryogenic parts of the QUBIC detection chain; the TD array of 64$+$64 back-to-back horns interfaced with their switches; the integrated 1K box; cryostat assembly of the radiation shields.}  
\label{fig:3}
\end{center}
\end{figure}

QUBIC experiment can be considered a Fizeau interferometer. In a Fizeau interferometer an array of apertures (the back-to-back horns) intercept the incoming radiation and each aperture illuminates the entire detector array. Each horn pair produces an interference (fringe) pattern defining a baseline and effectively selecting a Fourier mode in the sky. The sum of all baselines arising from the fields radiated from each horn pair defines the sensitivity to radiation coming from different directions of the sky (the synthesized beam, see Fig.~\ref{fig:2}, left). The synthetic image is then the convolution between the sky brightness and the synthesized beam of each pixel of the instrument ($B^p_{synth}$), so the expected signal at the detector $p$ is:
\begin{equation}
S_X(p) = \int X(\hat{n}) B^p_{synth}(\hat{n})d\hat{n} \label{eq3}
\end{equation}
being $X(\hat{n})$ the observed sky brightness whose polarized component will be modulated by the HWP rotation. 
The synthesized beam shape differs from those of standard imagers. The 150GHz beam has a primary maximum, with 0.54$^{\circ}$ Full Width at Half Maximum (FWHM), and secondary maxima offset by 8.5$^{\circ}$ from the central maximum. In addition, the effective synthesized beam depends on the frequency of the incident light demonstrating the spectral capability of an experiment as QUBIC (see Fig.~\ref{fig:2}, left). In Fig.~\ref{fig:2} we show the measured synthesized beam for one of the pixels of the Technological Demonstrator (TD) of the QUBIC instrument, compared to the theory (neglecting aberration effects). The QUBIC-TD consists of a reduced array of 64 back-to-back horns and one-quarter of the 150GHz TES array. It is currently being calibrated in the AstroParticle and Cosmology laboratory (APC), in Paris (see Fig.~\ref{fig:3} for assembling and parts examples). The TD uses the same cryostat and 1K box of the complete module, but has a reduced focal plane and horn array. 
\section{Operations}
QUBIC will be deployed in Argentina, at the Alto Chorrillos mountain site (4869 m a.s.l.) near San Antonio de los Cobres, in the Salta province at the end of 2019. The forecasted and monitored zenith optical depth in Alto Chorrillos at 210GHz, is $<$0.1 for 50\% of the year and $<$0.2 for 85\% of the year. Despite the fact that lower optical depths can be found in extreme observational sites like Dome C (see e.g. \cite{bat12}), the easier access and logistics have been considered to trade off performance and observation time, taking into account the fact that atmospheric emission is not polarized, and that an interferometer like QUBIC intrinsically rejects large-scale atmospheric gradients, which produce most of the atmospheric noise.

An alt-azimuthal mount will enable the pointing and the scanning. QUBIC will be able to rotate 360$^{\circ}$ in azimuth, from 30$^{\circ}$ to 70$^{\circ}$ in elevation, and 15$^{\circ}$ around the optical axis, with a pointing accuracy better than 20''. A forebaffle with 1m length and 14$^{\circ}$ aperture will be used to reduce radiation from unwanted sources by more than 20dB in directions between 20$^{\circ}$ and 40$^{\circ}$ away from the optical axis, and more than 40dB beyond. A ground shield will also be included to minimize the brightness contrast between the sky and the ground.

QUBIC will scan the sky at constant elevation over low-dust regions in the southern hemisphere, including the BICEP2 region (${\rm RA}=0^{\circ}$, ${\rm dec}=-57^{\circ}$) and the Planck clean field (${\rm RA}=8.7^{\circ}$, ${\rm dec}=-41.7^{\circ}$). When observed from Alto Chorrillos, the centers of these two regions change their elevation in the range 30$^{\circ}-$60$^{\circ}$ and 30$^{\circ}-$70$^{\circ}$ respectively, matching the allowed elevation range for the operation of the PTs. The instrument will typically scan in azimuth around the center of the selected field with a typical azimuth range of $\pm$15$^{\circ}$, and a speed $\sim$1$^{\circ}$/s. The elevation is updated after typically 10 scans to track the elevation of the center of the field. At the end of each scan the HWP is stepped by 15$^{\circ}$. Additionally, QUBIC is rotated in steps around its optical axis (bore-sight axis) to monitor systematic effects. This scanning strategy allows QUBIC to cover $\sim$1\% of the whole sky in 24h.  

During observations, QUBIC will alternate two operation modes: self-calibration and sky measurement. The self-calibration is used to select equivalent baselines (horn pairs that are at the same distance and with the same orientation in the array, that should produce the same fringe pattern on the focal planes) by using the RF switches.  We will use, for the self-calibration source, a microwave synthesizer able to radiate a typical power of few mW through a feedhorn with a well-known beam, and a low level of cross-polarization. It will generate tunable quasi-monochromatic signals and will be located in the far-field of the interferometer. Observing a partially polarized calibration artificial source during this self-calibration procedure, we can recover instrument parameters, including systematic effects, well constrained if sufficient integration time is used for the self-calibration, since the accuracy of recovered parameters is only limited by photon noise. As shown by Bigot-Sazy et al. \cite{self}, it is possible to reduce the impact of E-modes into B-modes leakage due to instrumental systematics, optimizing the observation time for each test baseline. This control of systematics, and the sensitivity of TES', allow QUBIC to push down the current limits on the tensor-to-scalar ratio, $r$, to $r<0.01$ at 95\%CL assuming two years of integration time over a sky patch similar to the BICEP2 one, an efficiency of 30\% related to the site quality, and 50\% duty cycle between observations and (self-)calibration, marginalizing over the foregrounds with the available frequency coverage. 

\section{Conclusions}
QUBIC is a novel experiment aiming to measure the B-modes of the polarization of the CMB. It combines the sensitivity of TES bolometers with the control of systematic effects typical of interferometers. The TD has already been integrated and is in the process of calibration, with very satisfactory preliminary results. We forecast the installation of the instrument in Argentina at the end of 2019. This will path the way for a new generation of instruments in the field of CMB polarimetry.

\begin{acknowledgements}
QUBIC is funded by the following agencies. France: ANR (Agence Nationale de la Recherche) 2012 and 2014, DIM-ACAV (Domaine d'Interet Majeur-Astronomie et Conditions d'Apparition de la Vie), CNRS/IN2P3 (Centre national de la recherche scientifique/Institut national de physique nucle\'aire et de physique des particules), CNRS/INSU (Centre  national de la recherche scientifique/Institut national de sciences de l'univers). Italy: CNR/PNRA (Consiglio Nazionale delle Ricerche/Programma Nazionale Ricerche in Antartide) until 2016, INFN (Istituto Nazionale di Fisica Nucleare) since 2017. Argentina: Secretar\'ia de Gobierno de Ciencia, Tecnolog\'ia e Innovaci\'on Productiva, Comisi\'on Nacional de Energ\'ia At\'omica, Consejo Nacional de Investigaciones Cient\'ificas y T\'ecnicas.
\end{acknowledgements}


\begin{thebibliography}{99}
\bibitem{planck16}
Planck collaboration, {\it A\&A} \textbf{594}, (2016), DOI:10.1051/0004-6361/201527101 
\bibitem{bicep18}
The Keck Array and BICEP2 Collaborations {\it  Phys. Rev. Lett.} \textbf{121}, 221301 (2018), DOI:10.1103/PhysRevLett.121.221301
\bibitem{osall18}
C. O'Sullivan et al. {\it SPIE Proc.} \textbf{10708}, 107082B, 14 (2018), DOI:10.1117/12.2313332
\bibitem{tar16}
A. Tartari et al. {\it JLTP} \textbf{184}, 3-4, 739-745 (2016), DOI:10.1007/s10909-015-1398-3
\bibitem{pia12}
M. Piat et al. {\it JLTP} \textbf{167}, 5-6, 872-878 (2012), DOI:10.1007/s10909-012-0522-x
\bibitem{bat11}
E.S. Battistelli et al. {\it  APP} \textbf{34}, 9, 705 (2011), DOI:10.1016/j.astropartphys.2011.01.012
\bibitem{may18}
A. May et al. {\it SPIE Proc.} \textbf{10708}, 107082V, 14 (2018), DOI:10.1117/12.2312085
\bibitem{del11}
F. Del Torto et al. {\it  JINST} \textbf{6}, 6009 (2011) DOI:10.1088/1748-0221/6/06/P06009
\bibitem{pia19}
M. Piat et al. {\it  These proceedings} 
\bibitem{mar19}
S. Marnieros et al. {\it  These proceedings} 
\bibitem{bat12}
E.Battistelli et al. {\it  MNRAS} \textbf{423}, 2, 1293 (2012) DOI:10.1111/j.1365-2966.2012.20951.x 
\bibitem{self}
M.A. Bigot-Sazy et al. {\it   A\&A} \textbf{550}, A59 (2013), DOI:10.1051/0004-6361/201220429 
\end{thebibliography}
\end{document}